\documentclass[a4paper,11pt]{article}
\pdfoutput=1 

\usepackage{jinstpub} 
\usepackage{subcaption}
\usepackage{lineno}
\usepackage[utf8]{inputenc}
\title{The semi-digital hadronic calorimeter (SDHCAL) for future leptonic colliders}

\author[a,1]{A. Pingault\note{Speaker}}

\affiliation[a]{Ghent University,\\Department of Physics and Astronomy,
Proeftuinstraat 86, B-9000 Gent, Belgium}

\emailAdd{antoine.pingault@ugent.be}

\abstract{The first technological SDHCAL prototype having been successfully tested, a new phase of R\&D, to validate completely the SDHCAL option for the International Linear Detector (ILD) project of the International Linear Collider (ILC), has started with the conception and the realisation of a new prototype. The new one is intended to host few but large active layers of the future SDHCAL. The new active layers, made of Glass Resistive Plate Chambers (GRPC) with sizes larger than 2~m$^{2}$ will be equipped with a new version of the electronic readout, fulfilling the requirements of the future ILD detector. The new GRPC are conceived to improve the homogeneity with a new gas distribution scheme. Finally the mechanical structure will be achieved using the electron beam welding technique. The progress realised will be presented and future steps will be discussed.}


\arxivnumber{1605.05075} 

\collaboration[c]{on behalf of the CALICE-SDHCAL groups}

\proceeding{XIII$^{\text{th}}$ Workshop on Resistive Plate Chambers and related detectors, \\
  Feb 22-26 2016\\
  Ghent University, Belgium}

\begin{document}
\maketitle
\flushbottom

\section{Introduction}
\label{sec:intro}

The CALICE-SDHCAL technological prototype has been successfully built and commissioned~\cite{Baulieu2015}. Its first published results~\cite{Buridon2016a} constitute a major step in the demonstration that highly-granular gaseous hadronic calorimeters are compatible with the requirements of the future ILC detectors in terms of efficiency, compactness and power consumption. A new prototype with fewer but larger layers with its dedicated mechanical structure, improved electronics and a new gas distribution system is in an advanced phase of development.

After a general description of the current SDHCAL prototype, the second section is dedicated to the presentation of the R\&D status for the next prototype. This includes discussions on the improvements made to the gas system, the electronics and the acquisition boards, the new mechanical structure developed to support the new chambers and a few other challenges.

\section{The CALICE-SDHCAL technological prototype}
\label{sec:currentPrototype}

The SDHCAL prototype hosts 48 GRPC with a size of 1$\times$1~m$^{2}$. Signals from the chambers are read out by 9216 pick-up pads of 1$\times$1~cm$^{2}$ each, corresponding to more than 442000 channels in the complete prototype. All the electronics is directly embedded inside the stainless steel structure of the layer. The prototype can work in power pulsed mode and take data triggerlessly as required for the ILC.
No cooling system is needed when running with a power pulsing cycle following the cycle from the ILC spill. For test beam purposes, where power pulsing is less efficient due to the spill cycle, a simple cooling system is used.

The prototype has been exposed to particles beams at CERN on several occasion during tests beam campaigns. Efficiency exceeding 90\% for the majority of the layers were obtained~\cite{Buridon2016a}. It also demonstrated, after calibration, a good energy resolution with a linear response (within $\pm~5\%$) over a large range of hadronic energies (5-80~GeV).

\section{Towards larger prototypes}
\label{sec:largerPrototype}

\begin{figure}[htbp]
\begin{subfigure}{.5\textwidth}
  \centering
  \includegraphics[width=.9\linewidth]{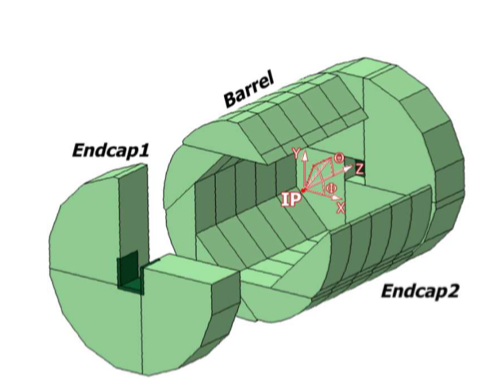}
  \caption{\label{subfig:videauBarrel} Full detector with barrel and endcaps.}
\end{subfigure}
\begin{subfigure}{.5\textwidth}
  \centering
  \includegraphics[width=.8\linewidth]{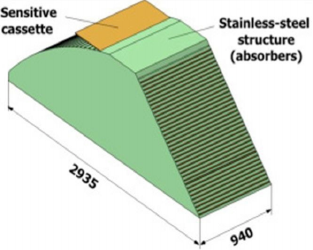}
  \caption{\label{subfig:videauModule} Single barrel module.}
\end{subfigure}
\caption{\label{fig:videauGeometry} Sketches of a HCAL using the Videau design~\cite{Behnke2010a}.}
\end{figure}

One of the geometry selected for the HCAL design of the ILD is the so-called Videau-geometry~\cite{Behnke2010a}. As can be seen on figure~\ref{fig:videauGeometry}, the particularity of this design is the variable size of the different layers constituting a single module. Whereas the width of 94~cm is the same for every active layer, their length is ranging from a few tens of centimetres to about 2.94~m. The first SDHCAL prototype proved that layers up to 1$\times$1~m$^{2}$ can be mass produced with good results, the next logical step is to demonstrate the same with chambers up to 3$\times$1~m$^{2}$.

The foreseen goal for the next couple of years is to achieve the construction and commissioning of a prototype containing few (3 to 4) active layers of $2\times1$~m$^{2}$. Given the good results of the $1\times1\times1.3$~m$^{3}$ prototype, one objective will be to keep the overall construction process wherever it is possible, capitalising on the know-how and equipment available.

\subsection{Gas distribution scheme}
\label{subsec:gas}
A proper homogeneity of the gas circulation inside the RPC volume is of the utmost importance to ensure high efficiency of detection and low noise.
When designing the gas distribution system one has to keep in mind that a requirement for the ILD is to have all the servicing of the detector on one side.

\begin{figure}[htbp]
\begin{subfigure}{.5\textwidth}
  \centering
  \includegraphics[width=.95\linewidth]{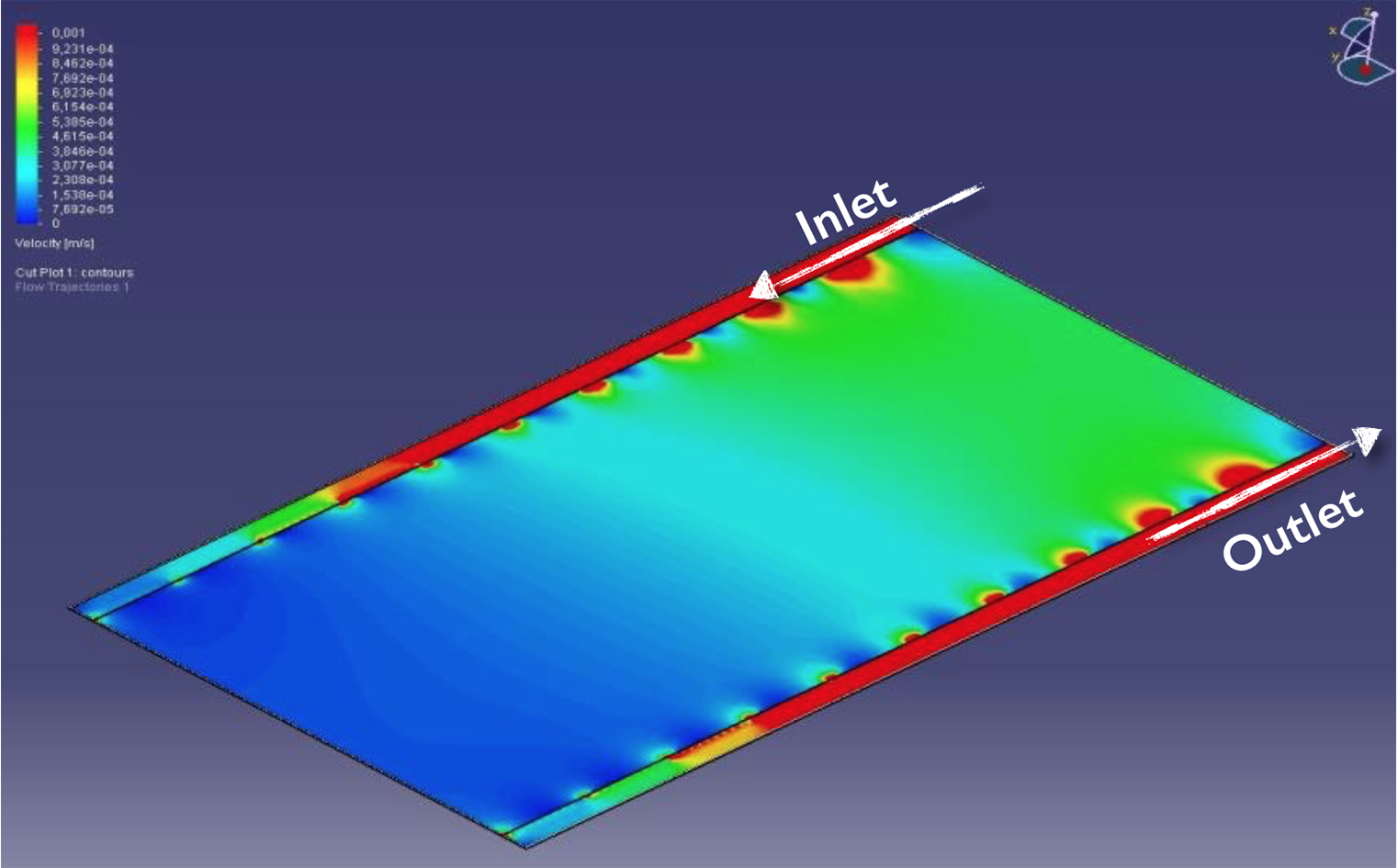}
  \caption{\label{subfig:oldGas} Current circulation system.}
\end{subfigure}
\begin{subfigure}{.5\textwidth}
  \centering
  \includegraphics[width=.95\linewidth]{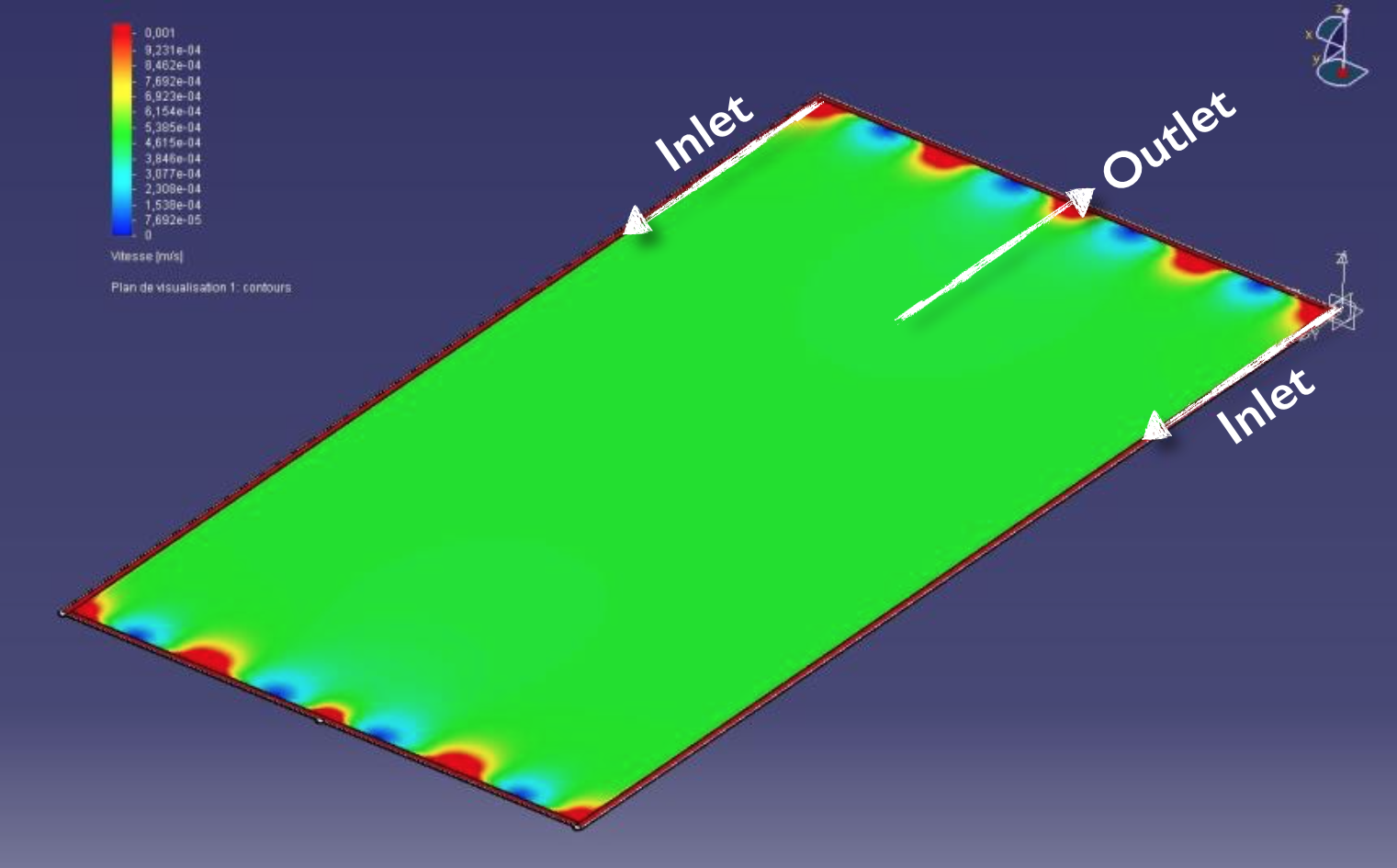}
  \caption{\label{subfig:newGas} New circulation system.}
\end{subfigure}
\caption{\label{fig:gasSystem} Gas circulation simulation in a $2\times1$~m$^{2}$ chamber.}
\end{figure}

The current set-up consists in a pair of inlet and outlet each at the opposite end of the servicing side. Inside the chamber along the frame following the inlet and outlet a small canal is perforated in several places to let the gas enter or exit the chamber.
Simulation studies conducted on the gas circulation scheme showed that this arrangement will be less efficient for chambers bigger than $1\times1$~m$^{2}$. A new scheme has been developed to improve the gas homogeneity.
In the new scheme the outlet is moved to the middle of the side and a second inlet is added in its place.
A view of the simulations for both systems in a $2\times1$~m$^{2}$ chamber can be seen in figure~\ref{fig:gasSystem}.

\subsection{Electronics}
\label{subsec:electronics}
The current electronics for each $1\times1$~m$^{2}$ chamber consist of three Detector InterFace (DIF) boards each attached to a slab of two daisy-chained Active Sensor Unit (ASU). An ASU namely hosts a Printed Circuit Board (PCB) with its 24 embedded HARDROC2~\cite{Callier2014} Application Specific Integrated Circuit (ASIC). Finally each ASIC can receive data from 64 channels. This adds up to 9216 channels connected to a matrix of $1\times1$~cm$^{2}$ copper pads in charge of reading out the electric signal from the RPC. A similar architecture with updated electronics will be used for larger prototypes, the total number of channels will then be tripled for the biggest active layer. In the following, improvements to each part of this electronic are presented.

\subsubsection{Detector Interface board}
\label{subsubsec:dif}
\begin{figure}[htbp]
  \centering
  \includegraphics[width=.85\textwidth]{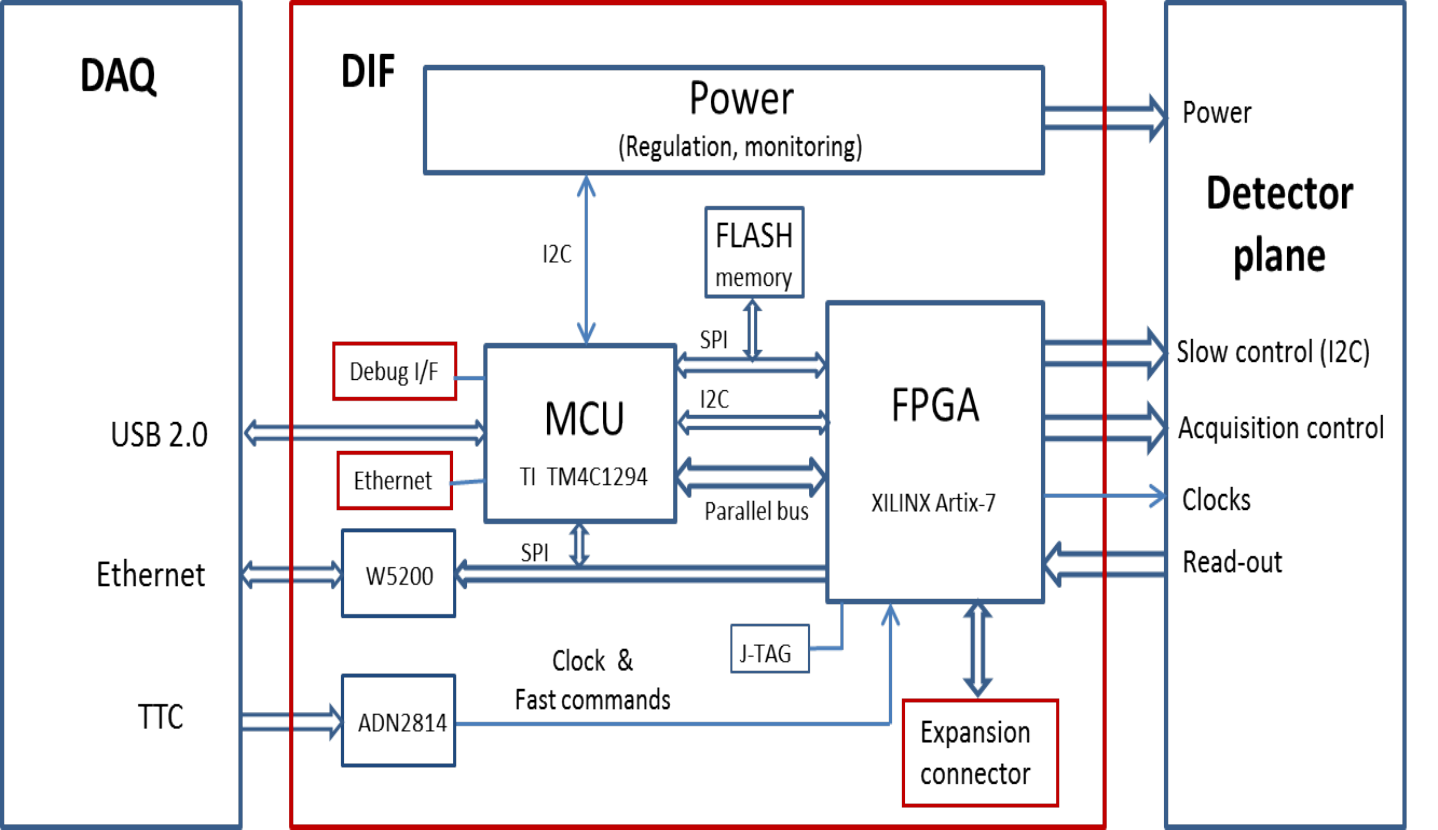}
  \caption{\label{fig:newDIF} Block diagram of the new SDHCAL DIF.}
\end{figure}

The DIF is used to link the Data Acquisition (DAQ) system to the ASICs. It will convey the DAQ commands (slow control, clock, etc.) to the ASICs on one way and the read out data from the ASICs to the DAQ system on the other way. A block diagram representing the functionalities of the new DIF board is shown in figure~\ref{fig:newDIF}.

The main goal here is to go from three boards to only one per chamber. This means that the new DIF will have to handle up to 432~HARDROC3 chips~\cite{Callier2014} for the biggest chamber instead of the 48~HARDROC2 chips currently. This will reduce consumption and simplify the cabling.

Other improvements on the functionalities were also done. Clocks signals are now sent through optical fibres using Timing, Trigger and Control signals (TTC)~\cite{Taylor1998} instead of the HDMI interface.
This is a much welcomed improvement as most of the synchronisation and cabling hassle on the current prototype is coming from these HDMI cables.
Passing of the slow control commands is done via 12 Inter-integrated Circuit (I2C) buses instead of shift registers giving the ability to control ASICs independently. Two of the old shift registers buses were kept for backup and redundancy.
Finally USB2 cables are replaced by Ethernet for data read out, increasing reliability and speed capability.

One last point still to be addressed concerns the connection to the ASU slab. Several constraints need to be met here:
\begin{itemize}
  \item The DIF length: the board has to be in a dedicated space outside the detection area. To meet the compactness requirement of the ILD, this dedicated area needs to be the smallest possible.
  \item The connector position between the DIF and the ASU.
  \item The high number of signals sent through this connection.
  \item The high current on the power pins: up to 30~A per plane.
\end{itemize}
One foreseen solution to meet all the constraints would be to integrate directly the DIF into a longer PCB.

\subsubsection{Active Sensor Unit}
\label{subsubsec:asu}
The ASU board as stated before is the main PCB board hosting the read out electronics. Keeping the same design for $2\times1$~m$^{2}$ layers would effectively double the number of ASU to ASU connection. Given that each connection adds a weak point to the electronics, one will want to reduce them to the minimum achievable. To this purpose, boards with a longitudinal size of 100~cm have been produced to host 48~ASICs.
This ensure a more viable design to scale to the biggest chamber.

\subsubsection{Application Specific Integrated Circuit}
\label{subsubsec:asic}
\begin{figure}[htbp]
  \centering
  \includegraphics[width=.45\textwidth]{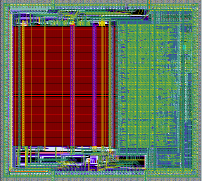}
  \caption{\label{fig:HR3} Schematics of the HARDROC3 chip.}
\end{figure}

A third generation of ROC chips has been developed by OMEGA\footnote{Laboratoire OMEGA - École Polytechnique, CNRS/IN2P3, Palaiseau, F-91128 France} and IPNLyon\footnote{Univ. Lyon, Université Lyon 1, CNRS/IN2P3, IPNL 4 rue E Fermi 69622, Villeurbanne CEDEX, France}. The new electronic will again use the HARDROC version, its schematic can be seen on figure~\ref{fig:HR3}.

The HARDROC3 chip has been updated to comply with the following ILD demands:
\begin{itemize}
  \item The 64 channels are now independent with zero suppression, and have current amplifier.
  \item The dynamic range is extended from 15pC to 50pC.
  \item Implementation of the I2C protocol for slow control parameters, this now gives the ability to control each asic individually.
  \item Integration of a fast clock generator inside the ASIC.
  \item HARDROC2 emulation for fall-back solution.
\end{itemize}

A few hundred chips were produced and tested. A yield of 83.3\% of them passed all the required tests, the majority of defect came from dead channels. Gain correction was successfully applied to effectively reduce the spread in the electronic response of the ASICs.

\subsection{Mechanical structure}
\label{subsec:mechanics}
Given the size of the chambers in development, a new mechanical structure is under construction by CIEMAT\footnote{CIEMAT, Centro de Investigaciones Energeticas, Medioambientales y Tecnologicas, Madrid, Spain}, with the collaboration of CERN services, in order to host them. The current structure for the $1\times1$~m$^{2}$ prototype is assembled with lateral spacers and staggered bolts. In order to reduce deformation and lateral dimensions the Electron Beam Welding (EBW) technique is under investigation for the assembly. Results using these more robust and less deforming welding techniques are encouraging.

One of the requirement for the structure is a thickness tolerance under 50~$\mu m$ and a planarity under 500~$\mu m$. This is already achieved for the $1\times1$~m$^{2}$ prototype but difficult, time consuming and expensive for longer length. To reduce this as much as possible, the roller levelling technique was chosen for this task. Planarity of less than 500~$\mu m$ over 3~m long layers was achieved~\cite{Fouz2015}.

\subsection{Other challenges}
\label{subsec:challenges}
One of the challenges not yet fully overcome is the construction of the cassette enclosing the chamber and its electronics. With the given design the total thickness of a chamber is of 11~mm, going to sizes of two to three metres may pose deformation issues of the cassettes leading to damage to the electrodes.
One obvious solution would be to increase the thickness of the cassette, meaning reducing the absorber in the mechanical structure in order to keep the same energy deposit in between layers. This would lead to a stiffer but heavier cassette. A full $1\times1$~m$^{2}$ chamber with 11~mm thickness is already weighting more than 70~kg, going to 3~m$^{2}$ chambers and augmenting the cassette weight might pose other handling issues, the mechanical structure will also need to be able to support this weight.
One of the major idea behind this detector is the fact that the cassette itself is part of the absorber, the other part being the mechanical structure. Thus to keep this concept, no light material like carbon fibre can be used to increase stiffness versus weight ratio.

The resistive coating is applied with the silk screen method using a liquid paint mix, then cured. The painting process might not pose trouble for bigger layer size but the curing step might as no production site is equipped with oven with the required size. Other coating methodology are under study to overcome this problem.

One recurrent issue in all the high energy physics experiments, and especially for gaseous detectors, is the heavy use of some gas, more often than not with a big green house factor. In order to reduce both the running costs and the impact on the environment, an ongoing effort, lead by the RPC-gas group at CERN, is focused on the development of recycling gas systems. This would lead to a gas renewal of the order of 5 to 10\% instead of 100\%. The 1~m$^{3}$ prototype is to be used for testing of this system. If conclusive, the next prototype will also be using this technology.

\section{Conclusion}
\label{sec:conclusion}

The efforts on the developments of the next CALICE-SDHCAL technological prototype are under way. The first of these prototypes will host few but larger layers without deteriorating the efficiency thanks to a renewed gas system. Lots of improvements has been made on the read out electronics notably giving the ability to control each ASIC individually and extending the dynamic range. The acquisition boards and the rest of the electronics have also been upgraded in order to keep up with the new features of the ASICs. The assembly of the mechanical structure with improved techniques is also in an advanced stage and is very encouraging.
Construction and testing of this prototype is scheduled to be done by next year. All challenges are not yet fully overcame but results are promising thanks to a fruitful collaboration inside and outside the SDHCAL group of the CALICE collaboration.

\acknowledgments
This work is conducted thanks to a fruitful collaboration between team from CIEMAT (Madrid, Spain), CERN (Geneva, Switzerland), GWNU (Gangneung, South Korea), IPNL (Lyon, France), LPC (Clermont-Ferrand, France), NCEPU (Beijing, China), OMEGA (Orsay, France),  UCL (Louvain, Belgium), UGent (Ghent, Belgium).

\end{document}